\documentclass[runningheads]{llncs}
\usepackage{orcidlink}
\usepackage{graphicx}
\usepackage{hyperref}
\usepackage[T1]{fontenc}
\usepackage{lmodern}
\usepackage{booktabs}
\usepackage{tabularx}   
\usepackage{multirow}
\usepackage{xcolor}
\usepackage{bbm}
\usepackage{cite}
\usepackage[normalem]{ulem}
\usepackage{enumitem}
\usepackage{amsmath}   
\usepackage{amssymb}   

\usepackage{makecell} 
\usepackage{array}    

\usepackage{marginnote}
\newcommand{\pageenlarge}[1]{\marginnote{}\enlargethispage{#1\baselineskip}}

\newcommand{\rd}[1]{#1}

\newcommand{\change}[1]{#1}
\newcommand{\qa}[1]{#1}
\newcommand{\gmd}[1]{#1}
\newcommand{\cam}[1]{#1}
\newcommand{\ioc}[1]{#1}

\begin{document}

\title{Temporal Conflicts in LLMs: Reproducibility Insights from Unifying DYNAMICQA and MULAN}
\titlerunning{Temporal Conflicts in LLMs: Unifying DYNAMICQA and MULAN}

\author{%
Ritajit Dey\inst{1}\orcidlink{0009-0006-3031-0722} \and
Iadh Ounis\inst{1}\orcidlink{0000-0003-4701-3223} \and
Graham McDonald\inst{1}\orcidlink{0000-0002-1266-5996} \and
Yashar Moshfeghi\inst{2}\orcidlink{0000-0003-4186-1088}}
\authorrunning{Dey et al.}

\institute{%
University of Glasgow, Glasgow, UK \\
\email{r.dey.1@research.gla.ac.uk} \\
\email{\{iadh.ounis,graham.mcdonald\}@glasgow.ac.uk}
\and
University of Strathclyde, Glasgow, UK \\
\email{yashar.moshfeghi@strath.ac.uk}
}

\maketitle

\begin{abstract}
Large Language Models (LLMs) often struggle with temporal fact conflicts due to outdated or evolving information in their training data. Two recent studies with accompanying datasets report opposite conclusions on whether external context can effectively resolve such conflicts. DYNAMICQA evaluates how effective external context is in shifting the model's output distribution, finding that temporal facts are more resistant to change. In contrast, MULAN examines how often external context changes memorised facts, concluding that temporal facts are easier to update. In this reproducibility paper, we first reproduce experiments from both benchmarks. We then reproduce the experiments of each study on the dataset of the other to investigate the source of their disagreement. To enable direct comparison of findings, we standardise both datasets to align with the evaluation settings of each study. Importantly, using an LLM, we synthetically generate realistic natural language contexts to replace MULAN’s programmatically constructed statements when reproducing the findings of DYNAMICQA. Our analysis reveals strong dataset dependence: MULAN’s findings generalise under both methodological frameworks, whereas applying MULAN’s evaluation to DYNAMICQA yields mixed outcomes. Finally, while the original studies only considered 7B LLMs, we reproduce these experiments across LLMs of varying sizes, revealing how \cam{model size} influences the encoding and updating of temporal facts. Our results highlight how dataset design, evaluation metrics, and model size shape LLM behaviour in the presence of temporal knowledge conflicts. \cam{Our \ioc{source} code and data are publicly available at \url{https://github.com/terrierteam/temporal_conflicts}.}
\end{abstract}

\section{Introduction}
\pageenlarge 1
\looseness -1 Large Language Models (LLMs) encode factual knowledge in their parameters, but inconsistencies in pretraining data often lead to \rd{knowledge} conflicts. A key source \rd{of knowledge conflicts} is temporal facts, where information changes over time, causing models to retain contradictory knowledge \cite{marjanovicDYNAMICQATracingInternal2024, suConflictBankBenchmarkEvaluating2024, xuKnowledgeConflictsLLMs2024a}. DYNAMICQA~\cite{marjanovicDYNAMICQATracingInternal2024} investigated these \rd{conflicts} \rd{(called intra-memory conflicts) in the context of a \qa{Question Answering (QA) task}}, by defining a broader category of dynamic facts, encompassing both temporal facts \rd{(changing with time, e.g., ``Who is the president of the US?'') and disputed facts (changing with viewpoint, e.g., ``What ethnic group is most commonly affected by lactose intolerance?'')}. To do this, \rd{Marjanovic et al. \cite{marjanovicDYNAMICQATracingInternal2024}} constructed a \rd{QA} dataset and showed that LLMs are less likely to be persuaded by external context, \qa{namely} Wikipedia passages paired with each fact, when dealing with dynamic facts compared to static ones. This finding raises concerns about the effectiveness of Retrieval-Augmented Generation (RAG), \qa{a typical approach} used to refresh or correct outdated knowledge~\cite{vuFreshLLMsRefreshingLarge2023}.

\looseness -1 In contrast, \qa{Fierro et al.~\cite{fierroMuLanStudyFact2024}}, \rd{devised a fact-completion benchmark, \qa{called MULAN}, \qa{in a study} of LLM behaviour on temporal facts. In terms of model performance, confidence, awareness and updateability \cam{(i.e., receptiveness of the LLM to knowledge updates)}}, they \qa{reported} the opposite pattern: temporal facts appear easier to update than static ones. Importantly, the two studies differ in how they define the effective use of external context: DYNAMICQA quantifies the extent to which context shifts the model’s output distribution (\textit{Coherent Persuasion score}), whereas MULAN measures the proportion of cases in which context changes a memorised answer (\textit{update success}). \rd{\qa{In addition}, unlike the natural-style passages of DYNAMICQA, MULAN uses simple programmatically constructed statements as external contexts.} These conflicting results reveal a lack of consensus on how \rd{receptive} LLMs \rd{ are to knowledge updates in the context of temporal facts}, motivating our reproducibility study.

\pageenlarge 1
\looseness -1 To investigate the source of these conflicting results, we reproduce each study’s evaluation framework on the other’s dataset. \qa{Moreover, when} applying DYNAMICQA’s Coherent Persuasion evaluation to MULAN, we replace \change{MULAN's} programmatically constructed statements with \cam{realistic} contexts generated by an LLM, inspired by CONFLICTBANK \cite{suConflictBankBenchmarkEvaluating2024}. Conversely, we apply MULAN’s update-success evaluation to DYNAMICQA, while adapting DYNAMICQA's question–answer format into MULAN's fact-completion style. This enables direct comparison of both frameworks. Moreover, DYNAMICQA \cite{marjanovicDYNAMICQATracingInternal2024} used the same relations (e.g., “capital of”) for both temporal and static facts and Wikipedia edit counts to estimate temporality, resulting in many temporal facts remaining unchanged. DYNAMICQA and MULAN also \qa{restricted} their experiments to 7B LLMs. \qa{In this reproducibility paper, we} extend the analysis to the Gemma 3 family (1B, 4B, 12B) \cite{teamGemma3Technical2025} to \qa{evaluate} whether findings hold across model \qa{sizes}. Using a single LLM family allows us to attribute differences more \rd{clearly} to \cam{model size} rather than architectural or training \qa{confounding variables} \cite{bidermanPythiaSuiteAnalyzing2023}.

\rd{\qa{To summarise,} our contributions in this paper are \qa{threefold}: (i) We conduct a systematic reproducibility study of the results reported in \qa{the} DYNAMICQA and MULAN \qa{studies}; (ii) We examine how dataset design and evaluation methodology influence conclusions about the updateability of temporal facts, \qa{highlight} how these factors contribute to the divergent findings of the DYNAMICQA and MULAN \qa{studies};
(iii) We analyse the effect of model \cam{size} on retaining and updating temporal facts.}

\section{Related Work}

\rd{In the following, we review related work on methods for updating factual knowledge in LLMs, and the factors that influence their success, situating our study within this broader literature.}

\noindent\rd{
\textbf{Knowledge Updates in LLMs.} Efforts to update factual knowledge in LLMs follow two main paths. One \qa{approach} is retrieval-augmented generation (RAG)~\cite{lewisRetrievalaugmentedGenerationKnowledgeintensive2020}, \qa{in which} the model conditions on retrieved evidence \qa{instead of updating} its parameters.
\rd{The other approach is parametric editing~\cite{decaoEditingFactualKnowledge2021, mitchellMemoryBasedModelEditing2022, mengLocatingEditingFactual2022a}, which modifies the model parameters to update the internal representation of a fact}.
However, Cohen et al.~\cite{cohenEvaluatingRippleEffects2024a} \qa{found} that in-context editing, \qa{consisting in} simply providing corrective information in the prompt, often yields more reliable updates than editing the parameters directly.}
\rd{Fierro et al.~\cite{fierroMuLanStudyFact2024} \qa{used} this in-context editing method \qa{in MULAN} to examine the receptiveness of LLMs to knowledge updates in the specific case of temporal facts.}

\noindent\rd{
\textbf{Factors Affecting Knowledge Updates.} Several studies \qa{have investigated} the factors that influence whether a knowledge update is successful. \qa{They showed} that LLMs often ignore context that contradicts their stored knowledge~\cite{longpreEntityBasedKnowledgeConflicts2021}, and that the receptiveness of the LLM to the external context depends on whether the context aligns with prior beliefs~\cite{xieAdaptiveChameleonStubborn2024} or is presented in a logically coherent form~\cite{xuEarthFlatBecause2024a}.
Li et al.~\cite{liInvestigatingContextFaithfulness2025} found that the strength of an LLM's memory for a given fact had a negative effect on the receptiveness of the LLM to external context.}

\pageenlarge 1
\rd{ \qa{Prior work has also demonstrated that} conflicting information in an LLM's pretraining data can lead to intra-memory conflicts, \qa{often due to} outdated or evolving facts~\cite{xuKnowledgeConflictsLLMs2024a, marjanovicDYNAMICQATracingInternal2024, suConflictBankBenchmarkEvaluating2024}.
Marjanovic et al.~\cite{marjanovicDYNAMICQATracingInternal2024} introduced the DYNAMICQA benchmark, \cam{a question-answering (QA) dataset}, to investigate \rd{the effect of} intra-memory conflicts \rd{on knowledge updates}, focusing on ``dynamic'' facts that change with time (temporal) or according to viewpoint (disputed). The \qa{DYNAMICQA} benchmark is built from Wikipedia edit histories\cam{, with each QA pair (\ioc{e.g.,} Q: ``Who is the screenwriter for The Shining'', A: ``Stanley Kubrick'') being accompanied by a natural passage as context}. Marjanovic et al.~\cite{marjanovicDYNAMICQATracingInternal2024} introduced the Coherent Persuasion score, which measures the extent to which conflicting context shifts the model’s output distribution. Using the calculated scores, \qa{they} found \qa{that} dynamic facts, \qa{including the} temporal ones, \qa{are} more resistant to updating.
However, Fierro et al.~\cite{fierroMuLanStudyFact2024} constructed a benchmark (MULAN) to evaluate LLM behaviour for temporal facts \cam{using fact-completion statements (e.g., ``The screenwriter for The Shining is...'')}, and reported the opposite conclusion: temporal facts are easier to update than \qa{the} static ones via external context. 
Fierro et al.~\cite{fierroMuLanStudyFact2024} collected temporal facts using Wikidata (subject, relation, object) triplets, which they paired with \cam{simple} programmatically constructed statements as external context (\cam{  e.g., ``Imagine that the screenwriter for [X] is [Y]. Then the screenwriter for [X] is...''}). \ioc{The process follows}\cam{~\cite{cohenEvaluatingRippleEffects2024a}, which \ioc{used} such statements as ``in-context knowledge edits''}. Fierro et al. evaluated the LLM's receptiveness to external context in terms of update success rate, \qa{that is} the proportion of cases in which external context changes a memorised answer.
Other contemporaneous work, such as CONFLICTBANK~\cite{suConflictBankBenchmarkEvaluating2024}, also examined knowledge updates under intra-memory conflict, but does so by synthetically injecting conflicting facts through continued pretraining rather than studying naturally occurring temporal changes.
\cam{Despite their shared focus, DYNAMICQA and MULAN differ in both \cam{their} dataset design and evaluation framework. DYNAMICQA, defines \cam{temporality} through \cam{Wikipedia} edit histories and measures distributional shift via \ioc{a} Coherent Persuasion score. In contrast, MULAN defines temporality through \cam{structured} relations \cam{extracted from Wikidata that are time-contingent and admit multiple possible values over time.} MULAN measures receptiveness to external context via update success.} These differences may explain their contradictory findings on updateability. Our work revisits \cam{the conflicting} conclusions from both studies} by aligning the datasets \ioc{used in} both studies and reproducing the methods of each study on the dataset of the other, to ascertain where the \cam{differences} arise.
Furthermore, DYNAMICQA and MULAN use only 7B LLMs in their experiments. We extend the scope of the evaluation to LLMs of different sizes to investigate the effect of model size on the findings of both studies. 

\section{Methodology}
\pageenlarge 1
This section outlines our experimental design, describing the three phases of our study, the methods reproduced from Marjanovic et al.~\cite{marjanovicDYNAMICQATracingInternal2024} \qa{(DYNAMICQA)} and Fierro et al.~\cite{fierroMuLanStudyFact2024} \qa{(MULAN)}, and the adaptations required to apply \rd{the evaluation methodology of each study across the DYNAMICQA and MULAN benchmarks}.

\looseness -1 \rd{First, we reproduce key evaluation methods from DYNAMICQA (Coherent Persuasion score, Semantic Entropy) and MULAN (Update Success Rate) with their original LLMs and datasets. Coherent Persuasion score and Update Success assess how well external context updates parametric knowledge, while Semantic Entropy evaluates output uncertainty. Following the original studies, we also report LLM performance across fact types and confidence estimates (via first-token probability) for MULAN.}

\looseness -1 \noindent\textbf{Coherent Persuasion Score.} \qa{In} DYNAMICQA, \qa{this measure} quantifies the semantic shift in an LLM’s output distribution when external context is provided. \rd{In a QA setting,} for each question $q$ and context $c_i$, two sets of answers are generated: 
$Y_{i;q}$ from the input without context $x_{i;q}$, and $Y_{i;q,c}$ from the input with context $x_{i;q,c}$. 
These generations are then clustered into semantically equivalent groups $G_{i;q}=\{g_1,\dots,g_R\}$ and $G_{i;q,c}=\{g_1,\dots,g_U\}$.
The CP score is defined as:
\begin{equation}
CP(c_i) = \frac{1}{R \times U} \sum_{r=1}^{R} \sum_{u=1}^{U} KL(p_{gr}, p_{gu})
\end{equation}
When computing $p_{g_r}$ and $p_{g_u}$, we note a discrepancy between the formula given in the paper and the one implemented in the released code. The paper states that the output probability distribution of $g_{r}$ and $g_{u}$ is obtained as \qa{follows}:
\begin{equation}
p_{g_w} = \frac{1}{W} \sum_{w=1}^{W} p_{y_w}
\end{equation}
\noindent where $W \in \{R, U\}$ and $p_{y_w}$ is the averaged softmax probability distribution across all tokens in the LLM's answer $a_w$. \qa{We note that this} definition is ambiguous: $w$ appears both as a subscript and a summation index, and $W$ is used to denote the number of groups while the summation is actually over answers within a group.
Inspecting the code, we find $p_{gr}$ or $p_{gu}$ to be \qa{as follows}:
\begin{equation}
p_{g_\ast} = \frac{1}{|g_\ast|} \sum_{j=1}^{|g_\ast|} p_{y j}, 
\quad g_\ast \in \{g_r, g_u\}.
\end{equation}
i.e., by averaging the probabilities of answers within each cluster.
We follow the version in the code.

\noindent\textbf{Update Success Rate.} \qa{In} MULAN, \qa{this is defined} as the proportion of cases in which external context changes a memorised fact.
Memorised facts are identified by selecting an equal number from each class of the dataset for which the LLM answers correctly with high confidence.
External contexts are constructed by replacing the ground-truth object of each fact with a randomly sampled alternative object from the same relation.
At inference time, the altered fact and the fact-completion query are combined into a fixed prompt: ``Imagine that <altered fact>. Then <fact completion query>''.
An update is considered successful if the LLM’s response exactly matches the replacement object.

\pageenlarge 1
\noindent\textbf{Semantic Entropy.} 
DYNAMICQA hypothesised that intra-memory conflicts increase output uncertainty, as competing answers lead to more dispersed distributions. They measured this using \textit{Semantic Entropy}~\cite{kuhnSemanticUncertaintyLinguistic2023}, estimated from diverse generations sampled at high temperature. 
Semantic Entropy clusters generations into semantically equivalent groups and computes entropy over these clusters, \qa{hence} uncertainty reflects competing meanings rather than surface forms.
\rd{We use the authors’ released code without modification.}

\rd{\qa{The next stage of our study consists in} reproducing the results of DYNAMICQA and MULAN on LLMs of varying size (Gemma 3 1B, 4B, and 12B), thereby extending beyond the original studies, which were restricted to 7B LLMs. We extend the released code of both benchmarks to support Gemma models, adding prompt construction and \qa{ensuring that} output processing \qa{is} consistent with the original implementations.}

\rd{Finally, the last stage of our study investigates how dataset design and evaluation methodology contribute to the diverging findings of both studies regarding the updateability of temporal facts. To this end, we apply the evaluation methodology of each study to the dataset of the other.
\rd{
To align with the temporal focus of MULAN, we exclude disputed facts when reproducing its methods on DYNAMICQA.} 
To do so, we first unify the formats of both datasets.
We convert the question templates in DYNAMICQA into the sentence-completion format found in MULAN, and vice versa.
Since both datasets were originally generated by instantiating a small set of fixed question/sentence templates with different subjects and objects, we extract the templates programmatically and manually adapt them into the format of the other dataset (``What was the director of [X]?'' becomes ``The director of [X] was'' and vice-versa).
Applying the evaluation methodology of DYNAMICQA on MULAN requires natural-style external contexts, whereas MULAN provides only programmatically constructed prompts. We generate Wikipedia-style contexts using an LLM (Llama-3.3-70B-Instruct), following CONFLICTBANK: we prompt the LLM to produce a supporting passage for each fact sentence. Outputs are filtered with an NLI model (\texttt{google/t5\_xxl\_true\_nli\_mixture}), retaining only those with entailment scores above $0.95$. \qa{We} further assess the quality of the generated passages by manually inspecting a random sample of 200 generations.
Following DYNAMICQA, contexts that conflict with the LLM's parametric knowledge are obtained by replacing the object in each snippet with a semantically similar entity retrieved using Wembedder~\cite{nielsenWembedderWikidataEntity2017}.
Because MULAN is much larger (49k samples vs. 5.7k in DYNAMICQA) and multi-sample methods such as Coherent Persuasion are computationally expensive, we subsample 2,500 instances per class with a fixed seed to create the converted dataset.}

\pageenlarge 1
\section{Experimental Setup}\label{sec:setup}
\rd{In this section, we outline our experimental setup, beginning with the \qa{used} datasets, 
followed by the evaluation metrics, then the LLMs with their decoding hyperparameters.} Our investigation \qa{aims to answer three} research questions:

\noindent \textbf{RQ1:} To what extent are the results reproducible when MULAN and DYNAMICQA are evaluated under their original experimental conditions?

\noindent \textbf{RQ2:} How does model \cam{size} influence an LLM’s ability to retain and update temporal facts when exposed to external context?

\noindent \textbf{RQ3:} How do dataset \rd{characteristics} and \cam{the used} evaluation methodology contribute to the \cam{conflicting} conclusions of DYNAMICQA and MULAN regarding the updateability of temporal facts?

\ \\
\noindent \textbf{Datasets:} \qa{In the following, we briefly describe the DYNAMICQA and MULAN datasets}. Table~\ref{tab:datasets} provides an overview of \qa{their statistics}.
\label{sec:datasets}

\begin{table}[h!]
\centering
\caption{Comparison of datasets with \ioc{their} task, size, source, and class categories.}
\label{tab:datasets}
\small
\begin{tabular}{p{2.2cm}|p{1.8cm}|p{1.7cm}|p{2.4cm}|p{3.3cm}}
\toprule
\textbf{Dataset} & \textbf{Task} & \textbf{\#samples} & \textbf{Source} & \textbf{Classes} \\
\midrule
DYNAMICQA & QA & 5,689 & Wikipedia facts & temporal, disputed, static \\
\midrule
MULAN & Sentence Completion & 49,142 & Wikidata triples & mutable, immutable-1, immutable-N \\
\bottomrule
\end{tabular}
\end{table}

\noindent\textbf{DYNAMICQA.} 
\qa{The dataset is a trivia QA benchmark for studying intra-memory conflicts, built from Wikipedia facts expressed as $(s, r, o)$ triplets across 16 relations. Each relation has a single template, and entries are categorised as static, temporal, or disputed, with temporality approximated via edit counts.} Each fact is paired with a Wikipedia snippet, in which the true answer is replaced by a substitute answer to obtain a context that conflicts with the LLM's parametric memory. \qa{A key limitation of DYNAMICQA is its use of the same relation set for static and temporal facts, relying solely on edit counts, thereby leading to misclassification of stable facts. Some relations (e.g., ``Who was the director of [X]?'') are inherently static, and apparent temporal variation often reflects disambiguation rather than true change. This motivates our re-evaluation using MULAN, which is explicitly designed for temporal facts.}

\

\noindent\textbf{MULAN.} The dataset is a fact-completion benchmark \qa{for studying} LLM behaviour on temporal facts. 
It is \qa{built} from $(s,r,o)$ triplets in Wikidata, \qa{spanning} 35 relations and five templates per relation. \qa{Facts are grouped into three classes: Immutable-1 (single static answer), Immutable-N (multiple static answers), and Mutable (time-sensitive with one valid answer per point in time).} 
Each object $o$ may also include multiple Wikidata aliases. In the original paper, both Immutable-1 and Immutable-N serve as controls for Mutable facts. For consistency, we treat Immutable-1 in MULAN as equivalent to \qa{DYNAMICQA's static class}, and Mutable as equivalent to the temporal class.

\ 

\pageenlarge 1
\noindent \textbf{Evaluation \qa{Metrics}:}
\label{sec:eval}
\noindent
In DYNAMICQA, accuracy is defined using Equations~\ref{eq:dqa_acc} and \ref{eq:dqa_acc2} for the with-context and without-context settings, respectively; 
\begin{equation}
\mathrm{acc} = \frac{1}{2N} \sum_{i=1}^{N} \Bigg(
\mathbbm{1}\!\left[ \mathrm{RougeL}\!\big(y_i^{o},\, a_o\big) > 0.3 \right] \;+\;
\mathbbm{1}\!\left[ \mathrm{RougeL}\!\big(y_i^{c},\, a_c\big) > 0.3 \right]
\Bigg)
\label{eq:dqa_acc}
\end{equation}

\begin{equation}
\mathrm{acc} = \frac{1}{N} \sum_{i=1}^{N} 
\mathbbm{1}\!\left[ \mathrm{RougeL}\!\big(y_i,\, a_o\big) > 0.3 \right]
\label{eq:dqa_acc2}
\end{equation}
\looseness -1where $y_i^o$ and $y_i^c$ (Equation~\ref{eq:dqa_acc}) denote the LLM outputs with the original and modified contexts, respectively, while $y_i$ (Equation~\ref{eq:dqa_acc2}) denotes the output without context. $a_o$ and $a_c$ are the original reference answer and the substitute answer (i.e., the target when the context is modified), and $N$ is the total number of examples.

However, we found that the released code \gmd{incorrectly} applies Equation~\ref{eq:dqa_acc} in \gmd{the} without-context setting, \gmd{instead of applying Equation~\ref{eq:dqa_acc2}}, meaning \gmd{that the} output, $y_i$, \gmd{in the without-context setting} is \gmd{additionally} compared to the incorrect \gmd{substitute} answer\gmd{, $a_c$}. \rd{In our experiments, we fix this by restricting the evaluation in the without-context setting to \gmd{comparing the output, $y_i$ to} the original \gmd{true} answer, following Equation~\ref{eq:dqa_acc2}.} Therefore, in Tables~\ref{tab:dqa_acc_main} and \ref{tab:dqa_acc_gemma} in Section~\ref{sec:results}, we report accuracy as measured by Equation~\ref{eq:dqa_acc2} for the without-context setting and Equation~\ref{eq:dqa_acc} for the with-context setting.

\rd{In addition, we reproduce all other measures introduced in DYNAMICQA, namely the Coherent Persuasion score, Semantic Entropy, and the proportions of stubborn and persuaded instances.
Stubborn instances are where the answer with context is identical to that without context.
Persuaded instances are where the context changes the LLM's answer to reflect the context.
}

\looseness -1 \rd{In MULAN,} predictions are evaluated using F1 against gold answers, including aliases. \qa{As is in the original MULAN study}, \qa{we also} report first-token confidence and Update Success Rate.

\noindent \textbf{Models and Hyperparameters:} We follow the released implementations of DYNAMICQA and MULAN unless otherwise noted, replicating their experimental settings as closely as possible.  For DYNAMICQA, we \qa{use} \rd{all LLMs evaluated in the original study:} Mistral-7B-Instruct-v0.1 \cite{jiangMistral7B2023}, Llama-2-7b-chat-hf \cite{touvronLlama2Open2023}, and Qwen2-7B-Instruct \cite{yangQwen2TechnicalReport2024}.
For MULAN, we use four of the six LLMs reported, \qa{namely} Llama 2 Base 7B \cite{touvronLlama2Open2023}, Llama 2 Chat 7B \cite{touvronLlama2Open2023}, Falcon Base 7B \cite{almazroueiFalconSeriesOpen2023}, and Falcon Instruct 7B, \rd{\qa{hence} omitting LLaMA 1 \cite{touvronLLaMAOpenEfficient2023} and Alpaca \cite{rohantaoriishaangulrajanitianyizhangyannduboisxuechenlicarlosguestrinpercyliangandtatsunorib.hashimotoStanfordAlpacaInstructionfollowing2023}}.
We also evaluate on the Gemma 3 family (1B, 4B, 12B)~ \cite{teamGemma3Technical2025} \rd{in order to assess the effect of model \cam{size} on the \change{retaining} and updateability of temporal facts.}

\rd{When evaluating accuracy in line with DYNAMICQA, following the original implementation, we use stochastic sampling with temperature 1.0, noting that this differs from the paper, which \qa{mentions} using greedy decoding}.
For \qa{the} Semantic Entropy and Coherent Persuasion \rd{measures}, consistent with the original implementation, we generate 10 samples per input at temperature 0.5, using DeBERTa-Large-MNLI as the NLI model. 
For \rd{MULAN-style evaluation, consistent with the released implementation} we use greedy decoding. 

\begin{table}[t]
\centering
\small
\caption{\qa{Accuracy} on DYNAMICQA and MULAN for Llama, Mistral, and Qwen. \cam{\ioc{Percentage differences between our results and those of the original study are shown in parentheses}. The large differences observed in the without-context accuracy values were caused by a correction to the DYNAMICQA evaluation code, which ensured that predicted answers were no longer compared with substitute answers (see Section~\ref{sec:setup}).}}
\label{tab:dqa_acc_main}
\resizebox{\textwidth}{!}{%
\begin{tabular}{l|ccc|ccc}
\hline
\textbf{Model} & \multicolumn{3}{c|}{\textbf{DYNAMICQA}} & \multicolumn{3}{c}{\textbf{MULAN}} \\
 & Static & Temporal & Disputed & Imm-1 & Imm-N & Mutable \\
\hline
Llama 2 Chat & 0.230 {\scriptsize (76.4\%)} & 0.177 {\scriptsize (71.0\%)} & -- & 0.562 & 0.514 & 0.386 \\
Llama 2 Chat w/context & 0.848 {\scriptsize (0.0\%)} & 0.660 {\scriptsize (-0.2\%)} & 0.646 {\scriptsize (0.0\%)} & 0.978 & 0.815 & 0.938 \\
Mistral Instruct & 0.166 {\scriptsize (83.6\%)} & 0.120 {\scriptsize (66.7\%)} & -- & 0.451 & 0.361 & 0.234 \\
Mistral Instruct w/context & 0.761 {\scriptsize (-0.5\%)} & 0.566 {\scriptsize (-0.3\%)} & 0.625 {\scriptsize (0.0\%)} & 0.926 & 0.787 & 0.884 \\
Qwen2 Instruct & 0.212 {\scriptsize (70.7\%)} & 0.157 {\scriptsize (81.0\%)} & -- & 0.543 & 0.429 & 0.334 \\
Qwen2 Instruct w/context & 0.685 {\scriptsize (0.5\%)} & 0.508 {\scriptsize (0.8\%)} & 0.591 {\scriptsize (-0.4\%)} & 0.899 & 0.752 & 0.800 \\
\hline
\end{tabular}%
}
\end{table}

\begin{table}[t]
\centering
\caption{Coherent Persuasion Score on DYNAMICQA and MULAN. \cam{\ioc{Percentage differences between our results and those of the original study are shown in parentheses}.}}
\label{tab:cp_score_main}
\begin{tabular}{l|ccc|ccc}
\toprule
\textbf{Models} & \multicolumn{3}{c|}{\textbf{DYNAMICQA}} & \multicolumn{3}{c}{\textbf{MULAN}} \\
\cmidrule(lr){2-4} \cmidrule(lr){5-7}
 & Static & Temporal & Disputed & Imm-1 & Imm-N & Mutable \\
\midrule
Llama 2 Chat & 6.856 {\scriptsize (-0.2\%)} & 6.612 {\scriptsize (0.3\%)} & 5.589 {\scriptsize (-0.3\%)} & 5.952 & 5.121 & 7.121 \\
Mistral Instruct & 5.843 {\scriptsize (-0.2\%)} & 5.656 {\scriptsize (0.4\%)} & 4.061 {\scriptsize (-1.3\%)} & 4.310 & 4.556 & 5.944 \\
Qwen2 Instruct & 3.936 {\scriptsize (-5.3\%)} & 3.381 {\scriptsize (-14.5\%)} & 3.418 {\scriptsize (0.7\%)} & 4.386 & 4.067 & 4.903 \\
\bottomrule
\end{tabular}
\end{table}
\vspace{-0.5em}

\begin{table}[t]
\caption{Semantic Entropy on DYNAMICQA. \cam{\ioc{Percentage differences between our results and those of the original study are shown in parentheses}.}}
\label{tab:se_dqa}
\centering
\begin{tabular}{lccc}
\toprule
\textbf{Models} & \textbf{Static} & \textbf{Temporal} & \textbf{Disputed} \\
\midrule
Llama 2            & 15.16 {\scriptsize (-14.4\%)} & 17.31 {\scriptsize (-0.2\%)} & 17.94 {\scriptsize (-5.4\%)} \\
Llama 2 w/context  & 15.13 {\scriptsize (-2.8\%)}  & 14.81 {\scriptsize (-3.8\%)} & 16.30 {\scriptsize (-1.7\%)} \\
Mistral            & 11.15 {\scriptsize (-3.0\%)}  & 11.32 {\scriptsize (-3.6\%)} & 11.96 {\scriptsize (-3.7\%)} \\
Mistral w/context  & 11.52 {\scriptsize (-2.0\%)}  & 10.62 {\scriptsize (-1.4\%)} & 11.60 {\scriptsize (-0.5\%)} \\
Qwen2              & 10.39 {\scriptsize (-0.4\%)}  & 11.04 {\scriptsize (-0.5\%)} & 10.31 {\scriptsize (-0.8\%)} \\
Qwen2 w/context    & 10.39 {\scriptsize (-1.9\%)}  & 10.30 {\scriptsize (-2.1\%)} & 10.92 {\scriptsize (-0.4\%)} \\
\bottomrule
\end{tabular}
\end{table}

\begin{table*}[h]
\centering
\caption{
\qa{Stubborn and persuaded instance percentages by fact type in DYNAMICQA.}
}
\label{tab:stubborn_persuaded_dqa}
\begin{tabular}{l|ccc|ccc}
\hline
\textbf{Models} & \multicolumn{3}{c|}{\textbf{\% of Stubborn Instances}} & \multicolumn{3}{c}{\textbf{\% of Persuaded Instances}} \\
 & Static & Temporal & Disputed & Static & Temporal & Disputed \\
\hline
Llama 2 Chat   & 3.60\% & 7.43\% & 9.08\% & 72.52\% & 57.70\% & 51.30\% \\
Mistral Instruct & 3.46\% & 5.57\% & 7.93\% & 68.50\% & 51.12\% & 49.06\% \\
Qwen2 Instruct   & 4.40\% & 5.35\% & 6.34\% & 59.06\% & 44.01\% & 47.98\% \\
Gemma 3 1B     & 4.78\% & 6.71\% & 7.35\% & 70.90\% & 55.63\% & 54.61\% \\
Gemma 3 4B     & 3.82\% & 6.11\% & 6.48\% & 72.10\% & 54.19\% & 59.08\% \\
Gemma 3 12B    & 3.96\% & 5.75\% & 6.12\% & 74.48\% & 59.50\% & 60.23\% \\
\hline
\end{tabular}
\vspace{-0.2em}
\end{table*}

\begin{table}[t]
\centering
\caption{Average F1 score and confidence of LLMs on MULAN. \cam{\ioc{ Percentage differences} between our results and those of the original study \ioc{are shown in parentheses} (for LLMs \ioc{included} in the original study only).}}
\label{tab:mulan_performance}
\resizebox{\linewidth}{!}{%
\begin{tabular}{l|cc|cc|cc}
\toprule
\textbf{Models} & \multicolumn{2}{c|}{\textbf{Imm-1}} & \multicolumn{2}{c|}{\textbf{Imm-N}} & \multicolumn{2}{c}{\textbf{Mutable}} \\
 & F1 & Conf & F1 & Conf & F1 & Conf \\
\midrule
Llama-2         & 42.62 {\scriptsize (-27.9\%)} & 0.36 {\scriptsize (-34.1\%)} & 35.56 {\scriptsize (-29.9\%)} & 0.27 {\scriptsize (-34.5\%)} & 17.00 {\scriptsize (-36.1\%)} & 0.18 {\scriptsize (-36.0\%)} \\
Llama-2-Chat    & 46.78 {\scriptsize (-17.1\%)} & 0.76 {\scriptsize (-13.3\%)} & 42.24 {\scriptsize (-17.0\%)} & 0.67 {\scriptsize (-19.1\%)} & 27.60 {\scriptsize (-14.0\%)} & 0.62 {\scriptsize (-21.0\%)} \\
Falcon          & 39.24 {\scriptsize (-22.9\%)} & 0.34 {\scriptsize (-8.6\%)}  & 33.03 {\scriptsize (-26.9\%)} & 0.25 {\scriptsize (-3.8\%)}  & 12.30 {\scriptsize (-28.9\%)} & 0.17 {\scriptsize (-8.2\%)} \\
Falcon Instruct & 36.50 {\scriptsize (-24.3\%)} & 0.46 {\scriptsize (-3.5\%)}  & 36.11 {\scriptsize (-20.1\%)} & 0.37 {\scriptsize (-4.9\%)}  & 17.16 {\scriptsize (-28.5\%)} & 0.26 {\scriptsize (-7.6\%)} \\
Gemma 1B IT     & 29.43 & 0.67 & 28.27 & 0.66 & 11.88 & 0.53 \\
Gemma 4B IT     & 43.87 & 0.84 & 36.05 & 0.83 & 20.30 & 0.75 \\
Gemma 12B IT    & 51.23 & 0.87 & 41.91 & 0.78 & 25.85 & 0.77 \\
\bottomrule
\end{tabular}}
\vspace{-0.2em}
\end{table}

\begin{table}[t]
\centering
\caption{Update Success Rate (as \qa{per} MULAN) on \textbf{MULAN} and \textbf{DYNAMICQA}. \cam{\ioc{Percentage differences between our results and those of the original study are shown in parentheses}.}}
\label{tab:update_success_main}
\begin{tabular}{lccc|cc}
\toprule
\textbf{Models} & \multicolumn{3}{c|}{\textbf{MULAN}} & \multicolumn{2}{c}{\textbf{DYNAMICQA}} \\
\cmidrule(lr){2-4} \cmidrule(lr){5-6}
 & \textbf{Imm-1} & \textbf{Imm-N} & \textbf{Mutable} & \textbf{Static} & \textbf{Temporal} \\
\midrule
Llama-2           & 0.598 {\small (5.6\%)}  & 0.640 {\small (-2.1\%)} & 0.720 {\small (-0.8\%)} & 0.551 & 0.558 \\
Llama-2 Chat      & 0.387 {\small (-2.6\%)} & 0.400 {\small (1.8\%)}  & 0.524 {\small (-0.4\%)} & 0.443 & 0.266 \\
Falcon            & 0.353 {\small (-21.4\%)}& 0.382 {\small (0.1\%)}  & 0.654 {\small (7.3\%)}  & 0.394 & 0.269 \\
Falcon Instruct   & 0.298 {\small (-1.2\%)} & 0.321 {\small (-3.1\%)} & 0.554 {\small (4.9\%)}  & 0.531 & 0.199 \\
\bottomrule
\end{tabular}
\end{table}

\begin{table*}[t]
\centering
\small
\caption{Accuracy (as \qa{per} DYNAMICQA) of Gemma LLMs on \textbf{DYNAMICQA}.}
\label{tab:dqa_acc_gemma}
\begin{tabular}{l|ccc}
\hline
\textbf{Model} & \multicolumn{3}{c}{\textbf{DYNAMICQA}} \\
 & Static & Temporal & Disputed \\
\hline
Gemma 3 1B IT & 0.122 & 0.088 & -- \\
Gemma 3 1B IT w/context & 0.775 & 0.595 & 0.626 \\
Gemma 3 4B IT & 0.175 & 0.140 & -- \\
Gemma 3 4B IT w/context & 0.808 & 0.601 & 0.734 \\
Gemma 3 12B IT & 0.241 & 0.185 & -- \\
Gemma 3 12B IT w/context & 0.848 & 0.670 & 0.793 \\
\hline
\end{tabular}
\end{table*}

\begin{table}[t]
\centering
\caption{Coherent Persuasion score on DYNAMICQA and Update Success Rate on MULAN for \qa{the} Gemma models.}
\label{tab:gemma_combined}
\begin{tabular}{l|ccc|ccc}
\toprule
\textbf{Models} 
& \multicolumn{3}{c|}{\shortstack{\textbf{DYNAMICQA} \\ {\scriptsize (Coherent Persuasion Score)}}} 
& \multicolumn{3}{c}{\shortstack{\textbf{MULAN} \\ {\scriptsize (Update Success Rate)}}} \\
\cmidrule(lr){2-4} \cmidrule(lr){5-7}
& Static & Temporal & Disputed & Imm-1 & Imm-N & Mutable \\
\midrule
Gemma 1B IT   & 9.976 & 9.746 & 7.833 & 0.694 & 0.462 & 0.687 \\
Gemma 4B IT   & 14.715 & 14.031 & 13.320 & 0.377 & 0.244 & 0.474 \\
Gemma 12B IT  & 10.142 & 10.008 & 9.837 & 0.438 & 0.321 & 0.437 \\
\bottomrule
\end{tabular}
\vspace{-1em}
\end{table}

\begin{table}[t]
\centering
\caption{\cam{Percentage of stubborn and persuaded instances, and Coherent Persuasion scores (as defined in DYNAMICQA) on MULAN.}}
\label{tab:stubborn_pers_cp_mulan}
\begin{tabular}{lcccccc|ccc}
\toprule
\textbf{Model} & \multicolumn{3}{c}{\% Stubborn Instances} & \multicolumn{3}{c}{\% Persuaded Instances} & \multicolumn{3}{|c}{Coherent Persuasion} \\
\cmidrule(lr){2-4} \cmidrule(lr){5-7} \cmidrule(lr){8-10}
 & Imm & Imm-N & Mut & Imm & Imm-N & Mut & Imm & Imm-N & Mut \\
\midrule
Llama 2 Chat     
& 2.68\% & 19.14\% & 2.68\%
& 70.42\% & 55.68\% & 79.54\%
& 5.952 & 5.121 & 7.121 \\

Mistral Instruct 
& 4.06\% & 11.54\% & 2.12\%
& 66.76\% & 59.92\% & 80.70\%
& 4.310 & 4.556 & 5.944 \\

Qwen2 Instruct   
& 3.06\% & 16.38\% & 1.82\%
& 64.20\% & 53.50\% & 68.56\%
& 4.386 & 4.067 & 4.903 \\
\bottomrule
\end{tabular}
\end{table}

\begin{table}[ht]
\centering
\caption{MULAN Performance comparison on DYNAMICQA data.}
\label{tab:mulaneval_dqadata}
\begin{tabular}{lcccc}
\toprule
\multirow{2}{*}{Models} & \multicolumn{2}{c}{Static} & \multicolumn{2}{c}{Temporal} \\
\cmidrule(lr){2-3} \cmidrule(lr){4-5}
 & F1 & Conf & F1 & Conf \\
\midrule
Llama-2          & 20.17 & 0.21 & 16.69 & 0.23 \\
Llama-2 Chat     & 22.54 & 0.60 & 24.16 & 0.61 \\
Falcon           & 15.91 & 0.21 & 12.79 & 0.24 \\
Falcon Instruct    & 15.49 & 0.30 & 15.63 & 0.34 \\
\bottomrule
\end{tabular}
\vspace{-0.5cm}
\end{table}

\pageenlarge 1
\section{Results}
\label{sec:results}

\looseness -1 We present \qa{reproducibility} results under \qa{the} original \rd{experimental} conditions \c{for DYNAMICQA and MULAN} (RQ1), \qa{examine} the effect of model \cam{size} on \rd{LLMs' ability to recall and update temporal facts} (RQ2), and \qa{analyse} how dataset design and evaluation methodology \qa{\qa{contribute} to the \cam{conflicting}} conclusions of \qa{the} \qa{studies} (RQ3).

\

\noindent \textbf{RQ1. To what extent are the results reproducible when evaluated under their original experimental conditions?}

\rd{
\rd{We reproduce the results of both studies by evaluating their LLMs on the fact types in their datasets, following the measures reported in each study, to compare LLM behaviour across fact types.}
\rd{We reproduce DYNAMICQA by using the LLMs to evaluate accuracy (see Table~\ref{tab:dqa_acc_main}), Coherent Persuasion score (see Table~\ref{tab:cp_score_main}), and Semantic Entropy (see Table~\ref{tab:se_dqa}) on the different fact partitions of the DYNAMICQA dataset (static, temporal, and disputed)}.
\rd{Following DYNAMICQA, we do not report accuracy for the disputed class in the without-context setting.}
We also report the percentage of stubborn instances (cases during the evaluation where the LLM's answer with context was the same as without context), and the percentage of persuaded instances (cases where providing the LLM with context changes the output to match the context) in Table~\ref{tab:stubborn_persuaded_dqa}.
For each measure, we compare the obtained scores on the temporal class against the static class.
Note that the mutable class consists of temporal facts, and the immutable classes do not change with time.}
\rd{We reproduce MULAN by using the LLMs to evaluate average F1 (see Table~\ref{tab:mulan_performance}) and Update Success rate (see Table~\ref{tab:update_success_main}) on the different fact partitions of MULAN (mutable, immutable-1, and immutable-N)}.
For each measure, we compare the obtained scores on the mutable class against the immutable-1 and immutable-N classes.

\ \\
\indent \rd{Our results are consistent with \qa{those reported by} DYNAMICQA, with no deviations that would affect the original conclusions.
DYNAMICQA evaluates the accuracy of an LLM in each of the fact partitions, finding lower LLM accuracy on dynamic (temporal and disputed) facts than static facts. This is supported by our results in Table~\ref{eq:dqa_acc}.
DYNAMICQA finds a lower Coherent Persuasion score for temporal and disputed facts than for static facts, as supported by our results in Table~\ref{tab:cp_score_main}.
In line with DYNAMICQA, we find a lower percentage of persuaded instances for dynamic facts than static facts and a higher percentage of stubborn instances for dynamic facts, as \qa{shown} in Table~\ref{tab:stubborn_persuaded_dqa}.
Consistent with DYNAMICQA, from Table~\ref{tab:se_dqa} we observe no trend of dynamic facts having higher Semantic Entropy than static facts.}
\qa{For MULAN, by comparing the average F1 score and confidence for mutable facts against immutable-1 and immutable-N facts, we can reproduce the finding that LLMs show lower F1 scores and confidence on mutable facts (see Table~\ref{tab:mulan_performance}).} Finally, we confirm higher update success for mutable facts than immutable-1 and immutable-N facts, as seen in Table~\ref{tab:update_success_main}, consistent with MULAN’s claim that they are easier to update.

\pageenlarge 1

\ 

\noindent \textbf{RQ2. How does model \cam{size} influence an LLM’s ability to retain and update temporal facts when exposed to external context?}

We reproduce the experiments in RQ1 on the Gemma series of 1B, 4B, and 12B LLMs, in contrast to the original studies, which restricted their evaluation to 7B LLMs.
\cam{This analysis is intended \ioc{to provide} a controlled \ioc{check of} robustness \ioc{and generalisation} across \ioc{different} model \cam{sizes}.} 
\qa{For DYNAMICQA, Table~\ref{tab:dqa_acc_gemma} reports LLM accuracy by fact type, while Table~\ref{tab:gemma_combined} presents the Coherent Persuasion scores}. We also provide the percentage of stubborn and persuaded instances in Table~\ref{tab:stubborn_persuaded_dqa}.
We do not evaluate Semantic Entropy when reproducing DYNAMICQA on the Gemma LLMs, \qa{since} the original study and our reproduced results already found no trend of higher Semantic Entropy for temporal facts.
For MULAN, we provide the results for the average F1 score and confidence of the LLMs for each fact type in Table~\ref{tab:mulan_performance} and the results for the Update Success Rate on each fact type in Table~\ref{tab:gemma_combined}.

Across \qa{the} Gemma models (1B, 4B, 12B), the trends reported in DYNAMICQA and MULAN generally hold, with one \cam{notable} exception: Update Success Rate in MULAN shows no consistent pattern of temporal facts being easier to update (Table~\ref{tab:gemma_combined}). This effect \cam{is observed only in Gemma 3 4B, whereas for both the 1B and 12B models}, mutable facts are updated less successfully than immutable-1 facts \cam{(\ioc{see} Section~\ref{sec:setup})}. \cam{Importantly, this results in a non-monotonic pattern: temporal facts are harder to update at both smaller and larger sizes, while the intermediate-size model shows the opposite behaviour. This finding suggests that conclusions about the updateability of temporal facts -- such as MULAN’s claim that they are generally easier to update~\cite{fierroMuLanStudyFact2024} -- are sensitive to model size and do not generalise even within a single model family.}
 
\

\pageenlarge 1
\noindent \textbf{RQ3. How do dataset characteristics and evaluation methodology contribute to the \cam{conflicting} conclusions of DYNAMICQA and MULAN regarding the updateability of temporal facts?}

\looseness -1 \rd{To examine the effect of dataset characteristics and evaluation methodology on the findings of DYNAMICQA and MULAN, we apply the evaluation methodology of either study to the dataset of the other.
We apply the Coherent Persuasion score of DYNAMICQA to the MULAN dataset (see Table~\ref{tab:cp_score_main}) as well as assess LLM accuracy on each fact type (Table~\ref{tab:dqa_acc_main}), after converting the fact-completion templates in MULAN into DYNAMICQA's QA format.
Similarly, we apply the Update Success measure of MULAN to the DYNAMICQA dataset (see Table~\ref{tab:mulaneval_dqadata}), \qa{and} assess F1 prediction score and LLM confidence on each fact type, after converting the question templates in DYNAMICQA into MULAN's fact-completion format.}

\pageenlarge 1
\looseness -1 We first observe that the findings of MULAN regarding mutable facts generalise to the evaluation methodology of DYNAMICQA.
From Table~\ref{tab:cp_score_main}, the mutable class has higher Coherent Persuasion scores across all evaluated LLMs.
Similarly, Table~\ref{tab:stubborn_pers_cp_mulan} shows that mutable facts yield the smallest proportion of stubborn instances and the highest proportion of persuaded instances.
General inference performance on the MULAN dataset (Table~\ref{tab:dqa_acc_main}) also mirrors MULAN’s original results (Table~\ref{tab:mulan_performance}): models perform worse on mutable than immutable facts.
Since MULAN does not consider with-context evaluation, the lower accuracy of immutable-N facts when context is provided (relative to immutable-1 and mutable) does not contradict its conclusions.
By contrast, the results of DYNAMICQA do not fully generalise to MULAN’s evaluation methodology.
Table~\ref{tab:update_success_main} shows that, aside from Llama-2, temporal facts have a lower update success rate, suggesting \qa{that} they are harder to update.
However, Table~\ref{tab:mulaneval_dqadata} \qa{shows} that only two of \qa{the} four models perform worse on temporal than \qa{on} static \qa{ones}, \qa{diverging} from DYNAMICQA’s original finding (Table~\ref{tab:dqa_acc_main}), \qa{which reported} that all models struggle more with temporal facts.
Moreover, LLMs consistently assign higher confidence to temporal facts than static ones.
Since higher confidence has been linked to stronger resistance to in-context updates \cite{liInvestigatingContextFaithfulness2025}, these results make it difficult to draw a firm conclusion that temporal facts are inherently harder to update.
The divergence between the two methodologies likely stems from both evaluation design and dataset construction.
Template sensitivity can affect LLM performance \cite{shaierComparingTemplatebasedTemplatefree2024}, and while MULAN uses multiple paraphrased templates, DYNAMICQA relies on a single one. \qa{We} attempted to mitigate this \qa{issue} by constructing five paraphrased templates for each question template in DYNAMICQA when converting into \qa{a} sentence-completion form.
Differences in how temporality is operationalised are also critical.
DYNAMICQA infers temporality via Wikipedia edit counts, without documenting quality checks or listing the specific fact versions used for the temporal class (only quality checks for the disputed class are documented).
Crucially, 1,369 of the 2,495 temporal facts have an edit count of just 2, only one more than the static facts. Such marginal differences may not consistently signal genuine temporal change and may instead reflect corrections or synonyms.
In contrast, MULAN determines temporality via relation type and object count in Wikidata, and provides code \qa{to generate} the dataset.
These discrepancies in dataset construction plausibly explain why general inference patterns and updateability judgments diverge across the two settings.
 
\pageenlarge 1
\section{Conclusions} 

We conducted a reproducibility study of two benchmarks on temporal knowledge conflicts in LLMs: DYNAMICQA and MULAN. We fully reproduced DYNAMICQA’s findings as well as MULAN’s results on temporal fact updateability, \cam{while correcting DYNAMICQA's evaluation code to ensure that \ioc{the} predicted answers were no longer compared with substitute answers}. Under their original experimental conditions, both studies’ claims were confirmed: DYNAMICQA found temporal facts harder to update, while MULAN found them easier \cam{to update}.
\cam{Furthermore, by applying each \ioc{of the benchmark's} evaluation framework to the other’s dataset and extending the analysis across different LLM sizes, we \ioc{showed} that these \cam{conflicting claims} do not generalise beyond their original settings. Differences in updateability between temporal and static facts \ioc{varied as to} whether temporal or static facts were easier to update, and \ioc{were} sensitive to dataset construction, definitions of temporality, evaluation metrics, and model size. Consequently, the conclusions drawn by both benchmarks are highly contingent on their specific experimental settings.}
\cam{Overall, neither benchmark alone \ioc{provides} a complete picture: \ioc{conclusions about temporal conflicts in LLMs are shaped by dataset construction, evaluation methodology, and model size.} \cam{Taken together, \ioc{our findings indicate that whether external context reliably resolves temporal knowledge conflicts in LLMs remains an open question.}}}

\subsubsection*{Acknowledgments}
\cam{This work was supported by the Engineering and Physical Sciences Research Council grant number EP/Y009800/1, through funding from Responsible Ai UK (KP0011).}

\subsubsection*{Disclosure of Interests}
\cam{The authors have no competing interests to declare that are relevant to the content of this article.}

\bibliographystyle{splncs04}
\bibliography{references}

@misc{almazroueiFalconSeriesOpen2023,
  author       = {Almazrouei, Ebtesam and others},
  title        = {The {Falcon} Series of Open Language Models},
  year         = {2023},
  howpublished = {arXiv preprint arXiv:2311.16867},
  url          = {https://arxiv.org/abs/2311.16867}
}

@misc{bidermanPythiaSuiteAnalyzing2023,
  author       = {Biderman, Stella and others},
  title        = {Pythia: A Suite for Analyzing Large Language Models Across Training and Scaling},
  year         = {2023},
  howpublished = {arXiv preprint arXiv:2304.01373},
  url          = {https://arxiv.org/abs/2304.01373}
}

@article{cohenEvaluatingRippleEffects2024a,
  author  = {Cohen, Roi and Biran, Eden and Yoran, Ori and Globerson, Amir and Geva, Mor},
  title   = {Evaluating the Ripple Effects of Knowledge Editing in Language Models},
  journal = {Transactions of the Association for Computational Linguistics},
  volume  = {12},
  pages   = {283--298},
  year    = {2024},
  doi     = {10.1162/tacl_a_00644}
}

@inproceedings{decaoEditingFactualKnowledge2021,
  author    = {De Cao, Nicola and Aziz, Wilker and Titov, Ivan},
  title     = {Editing Factual Knowledge in Language Models},
  booktitle = {Proceedings of the 2021 Conference on Empirical Methods in Natural Language Processing},
  pages     = {6491--6506},
  year      = {2021},
  doi       = {10.18653/v1/2021.emnlp-main.522}
}

@inproceedings{fierroMuLanStudyFact2024,
  author    = {Fierro, Constanza and Garneau, Nicolas and Bugliarello, Emanuele and Kementchedjhieva, Yova and S{\o}gaard, Anders},
  title     = {{MuLan}: A Study of Fact Mutability in Language Models},
  booktitle = {Proceedings of the 2024 Conference of the North American Chapter of the Association for Computational Linguistics: Human Language Technologies (Volume 2: Short Papers)},
  pages     = {762--771},
  year      = {2024},
  doi       = {10.18653/v1/2024.naacl-short.67}
}

@misc{jiangMistral7B2023,
  author       = {Jiang, Albert Q. and others},
  title        = {Mistral 7B},
  year         = {2023},
  howpublished = {arXiv preprint arXiv:2310.06825},
  url          = {https://arxiv.org/abs/2310.06825}
}

@inproceedings{kuhnSemanticUncertaintyLinguistic2023,
  author    = {Kuhn, Lorenz and Gal, Yarin and Farquhar, Sebastian},
  title     = {Semantic Uncertainty: Linguistic Invariances for Uncertainty Estimation in Natural Language Generation},
  booktitle = {Proceedings of the 11th International Conference on Learning Representations},
  year      = {2023},
  url       = {https://openreview.net/forum?id=VD-AYtP0dve}
}

@inproceedings{lewisRetrievalaugmentedGenerationKnowledgeintensive2020,
  author    = {Lewis, Patrick and Perez, Ethan and Piktus, Aleksandra and Petroni, Fabio and Karpukhin, Vladimir and Goyal, Naman and K{\"u}ttler, Heinrich and Lewis, Mike and Yih, Wen-tau and Rockt{\"a}schel, Tim and Riedel, Sebastian and Kiela, Douwe},
  title     = {Retrieval-Augmented Generation for Knowledge-Intensive NLP Tasks},
  booktitle = {Advances in Neural Information Processing Systems},
  volume    = {33},
  pages     = {9459--9474},
  year      = {2020},
  url       = {https://proceedings.neurips.cc/paper_files/paper/2020/file/6b493230205f780e1bc26945df7481e5-Paper.pdf}
}

@inproceedings{liInvestigatingContextFaithfulness2025,
  author    = {Li, Yuepei and Zhou, Kang and Qiao, Qiao and Nguyen, Bach and Wang, Qing and Li, Qi},
  title     = {Investigating Context Faithfulness in Large Language Models: The Roles of Memory Strength and Evidence Style},
  booktitle = {Findings of the Association for Computational Linguistics: ACL 2025},
  pages     = {4789--4807},
  year      = {2025},
  doi       = {10.18653/v1/2025.findings-acl.247}
}

@inproceedings{longpreEntityBasedKnowledgeConflicts2021,
  author    = {Longpre, Shayne and Perisetla, Kartik and Chen, Anthony and Ramesh, Nikhil and DuBois, Chris and Singh, Sameer},
  title     = {Entity-Based Knowledge Conflicts in Question Answering},
  booktitle = {Proceedings of the 2021 Conference on Empirical Methods in Natural Language Processing},
  pages     = {7052--7063},
  year      = {2021},
  doi       = {10.18653/v1/2021.emnlp-main.565}
}

@inproceedings{marjanovicDYNAMICQATracingInternal2024,
  author    = {Marjanovi{\'c}, Sara Vera and Yu, Haeun and Atanasova, Pepa and Maistro, Maria and Lioma, Christina and Augenstein, Isabelle},
  title     = {{DYNAMICQA}: Tracing Internal Knowledge Conflicts in Language Models},
  booktitle = {Findings of the Association for Computational Linguistics: EMNLP 2024},
  pages     = {14346--14360},
  year      = {2024},
  doi       = {10.18653/v1/2024.findings-emnlp.838}
}

@inproceedings{mengLocatingEditingFactual2022a,
  author    = {Meng, Kevin and Bau, David and Andonian, Alex and Belinkov, Yonatan},
  title     = {Locating and Editing Factual Associations in GPT},
  booktitle = {Advances in Neural Information Processing Systems},
  volume    = {35},
  pages     = {17359--17372},
  year      = {2022},
  url       = {https://proceedings.neurips.cc/paper_files/paper/2022/file/6f1d43d5a82a37e89b0665b33bf3a182-Paper-Conference.pdf}
}

@inproceedings{mitchellMemoryBasedModelEditing2022,
  author    = {Mitchell, Eric and Lin, Charles and Bosselut, Antoine and Manning, Christopher D. and Finn, Chelsea},
  title     = {Memory-Based Model Editing at Scale},
  booktitle = {Proceedings of the 39th International Conference on Machine Learning},
  pages     = {15817--15831},
  year      = {2022},
  url       = {https://proceedings.mlr.press/v162/mitchell22a.html}
}

@misc{nielsenWembedderWikidataEntity2017,
  author       = {Nielsen, Finn {\AA}rup},
  title        = {Wembedder: Wikidata Entity Embedding Web Service},
  year         = {2017},
  howpublished = {arXiv preprint arXiv:1710.04099},
  url          = {https://arxiv.org/abs/1710.04099}
}

@misc{rohantaoriishaangulrajanitianyizhangyannduboisxuechenlicarlosguestrinpercyliangandtatsunorib.hashimotoStanfordAlpacaInstructionfollowing2023,
  author = {Taori, Rohan and Gulrajani, Ishaan and Zhang, Tianyi and Dubois, Yann and Li, Xuechen and Guestrin, Carlos and Liang, Percy and Hashimoto, Tatsunori B.},
  title  = {{Stanford Alpaca}: An Instruction-Following {LLaMA} Model},
  year   = {2023},
  howpublished = {\url{https://github.com/tatsu-lab/stanford_alpaca}}
}

@inproceedings{shaierComparingTemplatebasedTemplatefree2024,
  author    = {Shaier, Sagi and Bennett, Kevin and Hunter, Lawrence and von der Wense, Katharina},
  title     = {Comparing Template-Based and Template-Free Language Model Probing},
  booktitle = {Proceedings of the 18th Conference of the European Chapter of the Association for Computational Linguistics (Volume 1: Long Papers)},
  pages     = {766--776},
  year      = {2024},
  doi       = {10.18653/v1/2024.eacl-long.46}
}

@inproceedings{suConflictBankBenchmarkEvaluating2024,
  author    = {Su, Zhaochen and Zhang, Jun and Qu, Xiaoye and Zhu, Tong and Li, Yanshu and Sun, Jiashuo and Li, Juntao and Zhang, Min and Cheng, Yu},
  title     = {{ConflictBank}: A Benchmark for Evaluating the Influence of Knowledge Conflicts in LLMs},
  booktitle = {Advances in Neural Information Processing Systems},
  volume    = {37},
  pages     = {103242--103268},
  year      = {2024},
  url       = {https://proceedings.neurips.cc/paper_files/paper/2024/file/baf4b960d118f838ad0b2c08247a9ebe-Paper-Datasets_and_Benchmarks_Track.pdf}
}

@misc{teamGemma3Technical2025,
  author       = {{Gemma Team} and Kamath, Aishwarya and others},
  title        = {Gemma 3 Technical Report},
  year         = {2025},
  howpublished = {arXiv preprint arXiv:2503.19786},
  url          = {https://arxiv.org/abs/2503.19786}
}

@misc{touvronLlama2Open2023,
  author       = {Touvron, Hugo and others},
  title        = {Llama 2: Open Foundation and Fine-Tuned Chat Models},
  year         = {2023},
  howpublished = {arXiv preprint arXiv:2307.09288},
  url          = {https://arxiv.org/abs/2307.09288}
}

@misc{touvronLLaMAOpenEfficient2023,
  author       = {Touvron, Hugo and others},
  title        = {{LLaMA}: Open and Efficient Foundation Language Models},
  year         = {2023},
  howpublished = {arXiv preprint arXiv:2302.13971},
  url          = {https://arxiv.org/abs/2302.13971}
}

@misc{vuFreshLLMsRefreshingLarge2023,
  author       = {Vu, Tu and Iyyer, Mohit and Wang, Xuezhi and Constant, Noah and Wei, Jerry and Wei, Jason and Tar, Chris and Sung, Yun-Hsuan and Zhou, Denny and Le, Quoc and Luong, Thang},
  title        = {{FreshLLMs}: Refreshing Large Language Models with Search Engine Augmentation},
  year         = {2023},
  howpublished = {arXiv preprint arXiv:2310.03214},
  url          = {https://arxiv.org/abs/2310.03214}
}

@misc{xieAdaptiveChameleonStubborn2024,
  author       = {Xie, Jian and Zhang, Kai and Chen, Jiangjie and Lou, Renze and Su, Yu},
  title        = {Adaptive Chameleon or Stubborn Sloth: Revealing the Behavior of Large Language Models in Knowledge Conflicts},
  year         = {2024},
  howpublished = {arXiv preprint arXiv:2305.13300},
  url          = {https://arxiv.org/abs/2305.13300}
}

@inproceedings{xuEarthFlatBecause2024a,
  author    = {Xu, Rongwu and Lin, Brian and Yang, Shujian and Zhang, Tianqi and Shi, Weiyan and Zhang, Tianwei and Fang, Zhixuan and Xu, Wei and Qiu, Han},
  title     = {The Earth Is Flat Because...: Investigating {LLMs'} Belief towards Misinformation via Persuasive Conversation},
  booktitle = {Proceedings of the 62nd Annual Meeting of the Association for Computational Linguistics (Volume 1: Long Papers)},
  pages     = {16259--16303},
  year      = {2024},
  doi       = {10.18653/v1/2024.acl-long.858}
}

@inproceedings{xuKnowledgeConflictsLLMs2024a,
  author    = {Xu, Rongwu and Qi, Zehan and Guo, Zhijiang and Wang, Cunxiang and Wang, Hongru and Zhang, Yue and Xu, Wei},
  title     = {Knowledge Conflicts for {LLMs}: A Survey},
  booktitle = {Proceedings of the 2024 Conference on Empirical Methods in Natural Language Processing},
  pages     = {8541--8565},
  year      = {2024},
  doi       = {10.18653/v1/2024.emnlp-main.486}
}

@misc{yangQwen2TechnicalReport2024,
  author       = {Yang, An and others},
  title        = {Qwen2 Technical Report},
  year         = {2024},
  howpublished = {arXiv preprint arXiv:2407.10671},
  url          = {https://arxiv.org/abs/2407.10671}
}

\end{document}